# Integration of fluorescence collection optics with a microfabricated surface electrode ion trap

G.R. Brady[‡], A. R. Ellis, D. L. Moehring, D. Stick, C. Highstrete, K. M. Fortier, M. G. Blain, R. A. Haltli, A. A. Cruz-Cabrera, R. D. Briggs, J. R. Wendt, T. R. Carter, S. Samora and S. A. Kemme

Sandia National Laboratories, P.O. Box 5800, Albuquerque, NM 87185-1082, USA

Email: grbrady@sandia.gov

**Abstract.** We have successfully demonstrated an integrated optical system for collecting the fluorescence from a trapped ion. The system, consisting of an array of transmissive, dielectric micro-optics and an optical fiber array, has been intimately incorporated into the ion-trapping chip without negatively impacting trapping performance. Epoxies, vacuum feedthrough, and optical component materials were carefully chosen so that they did not degrade the vacuum environment, and we have demonstrated light detection as well as ion trapping and shuttling behavior comparable to trapping chips without integrated optics, with no modification to the control voltages of the trapping chip.

[‡] Author to whom any correspondence should be addressed.

## 1. Introduction

The efficient collection of fluorescence photons from a single ion is an essential ingredient for trapped ion-based quantum information processing [1]. The time-resolved detection of emitted photons has allowed for >99.99% single-ion qubit readout fidelities where greater photon collection efficiencies lead to higher detection fidelities [2], [3]. Furthermore, the collection efficiency of individually emitted photons is critical for photon-mediated remote ion entanglement [4].

This work is a significant realization on a path previously proposed by Kielpinski, et al. [5] to achieve higher collection efficiencies and simultaneous point detection of ions via an optical system integrated with the trapping chip. We have designed and implemented a lens and fiber optic system that intimately integrates with the ion-trapping chip to access the largest possible portion of the ion's fluorescence, which is radiated into a full sphere, given the constraints imposed by the ion trapping configuration. The system also permits us to discriminate an ion's fluorescence from its neighbor's. This approach employs an array of high numerical aperture (NA) dielectric diffractive microlenses that efficiently and robustly transmits the ion fluorescence to an array of multi-mode optical fibers [6]. This integration is on a finer scale than has previously been demonstrated with macro, three-dimensional Paul traps whose dimensions are on the order of millimeters. Our approach functions complementarily and simultaneously to existing bulk optics that conventionally monitor ion presence, collect fluorescence and illuminate with laser light. Some advantages of our approach over other in-vacuum designs [7]-[9] are closer integration of the transmitting, dielectric optical element array to the trapped ion than was previously attempted, an ability to realize full fill-factor, low aberration optics, and built-in redundancy and location tolerance along the trap axis through the use of an array of optical elements without occluding any portion of the fluorescence. Other trap designs proposing [10] or demonstrating [11], [12] microscale optics for ion trapping typically collimate the fluorescence beam to relay it to the exterior of the chamber, while we couple directly into fibers attached to the ion-trapping chip.

The approach described here presents many potential risks to the trapping system. These were managed by careful consideration of the optical design form used and careful selection of materials that are compatible with the ultra-high vacuum (UHV) environment required by the trap. A major concern is that the close proximity of a dielectric lens to the ion trap might disrupt the electric fields to such an extent that the fundamental trapping and shuttling operations would be destabilized. Specifically, the introduction of a dielectric surface near an illuminating light source can give rise to excess micromotion and difficulties in reliably addressing ions with focused laser beams. This is thought to be a result of slowly drifting charges in the vicinity of trapped ions and on the dielectric [13]. Other researchers have considered this issue at the macro-trap scale 7, and at the micro-trap scale [8]. This micro-trap arrangement is similar to ours, but does not have the advantage of an optical element of any power to enhance collection efficiency.

We have detected light through the diffractive micro-optics/fiber system and will quantify our optical system efficiency measurement and the resulting fluorescence collection in a future publication.

In this paper we discuss the steps we have taken to design and successfully integrate collection optics with the trap chip and to mitigate risks associated with the introduction of the optics. We detail our design considerations in Sec. 2. In Sec. 3 we discuss the design of high numerical-aperture, transmissive diffractive micro-optics that collect ion fluorescence and couple it efficiently to an array of optical fibers, as well as a scheme for integrating the optics directly with a packaged ion-traping chip.  In Sec. 4 we describe the fabrication of the diffractive optics, the alignment assembly of the optics with the trap chip, and the assembly of the trap chip and optics in the vacuum chamber.  In Sec. 5 we describe our successful trapping and shuttling of ions using the chip with integrated optics and we conclude in Sec. 6.

## 2. Design Considerations

### 2.1. Integrating Optics with Trap Chip

It is only in considering the combined system aspects of the trap chip with integrated optics that an optimum, practical configuration is identified.  In Figure 1 we show a simplified drawing of the architecture assumed in this work.  An excitation beam from a single mode optical fiber is focused by a microlens onto a trapped ion.  The resulting fluorescence is captured by a collection microlens and coupled into a multimode optical fiber.  This is conceptually simple, but in reality there are many constraints on the optical design because of the necessity of integrating the optics with a functioning ion-trapping chip, such as that shown in schematic in Figure 2.  The geometry of the trap chip is tightly coupled to the trapping function and can be altered minimally or not at all for the purpose of integrating optics.

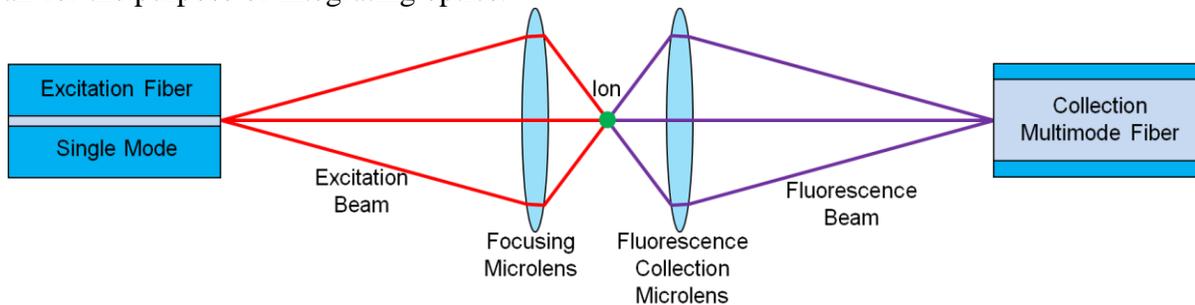

**Figure 1.**  Layout for exciting a trapped ion and collecting the resulting fluorescence.  The excitation and fluorescence may be temporally and spatially separated.

A key consideration is that the emission from a single ion is weak and into $4\pi$ steradians, so the collection optics should collect light over as large a solid angle as possible, i.e. the optics should be of high NA.  Unfortunately there are a number of competing considerations that limit the cone of light that can be collected.  The ion is held in the trap using electric fields so static charges on nearby dielectrics, such as our lens aperture, produce a competing electric field that can perturb the ion trapping field.  As a result, the lens may not be placed particularly close to the ion.  The cone size is also limited by a choice to place the fluorescence collection optics on the opposite side of the chip from the trapped ion, collecting light through the chip's loading slot.  This choice was made to simplify the incorporation of the trap with integrated optics into existing trapping arrangements and allow for the use of conventional excitation and collection optics.  The fluorescence is collected through a 100 μm wide slot in the trap electrodes.  This slot thus becomes the limiting aperture (or aperture stop) of the system.  Since its placement is determined entirely by the design of the trap chip, a degree of freedom commonly used in optical designs is not available. Rather, our optics must be designed to accommodate its placement.  The size of the slot and placement of the ion relative to it correspond to a solid angle of 0.45 Sr, 3.6% of a whole sphere, or a numerical aperture of 0.37.

Another consideration is that the lateral dimension of the lens aperture must be kept small to allow for dense integration of the lenses into arrays, enabling the detection of fluorescence from ions at multiple locations along the trap. This consideration requires that the lens be near the ion, in direct opposition to the desired placement to mitigate charging effects.

In addition, the absolute location of the trapped ion may not be well-controlled, requiring a design with loose positioning tolerances to accommodate these uncertainties. It is anticipated that the lateral position of the ion may vary by as much as ±10 μm, which corresponds to a field angle of as much as 5 degrees. The height of the ion above the trap electrodes (along the optical axis) may also vary significantly, requiring a design robust to focus errors.

Finally, the operating wavelength is defined by the wavelength of the fluorescent transition of interest, which is a function of which element is chosen for the trapped ion. This is often in the UV. In this work, we utilized the $4S_{1/2} - 4P_{1/2}$ transition in $Ca^+$ at 397nm. The short wavelength limits our choice of optical materials and complicates the fabrication of the diffractive optics due to the small features necessary.

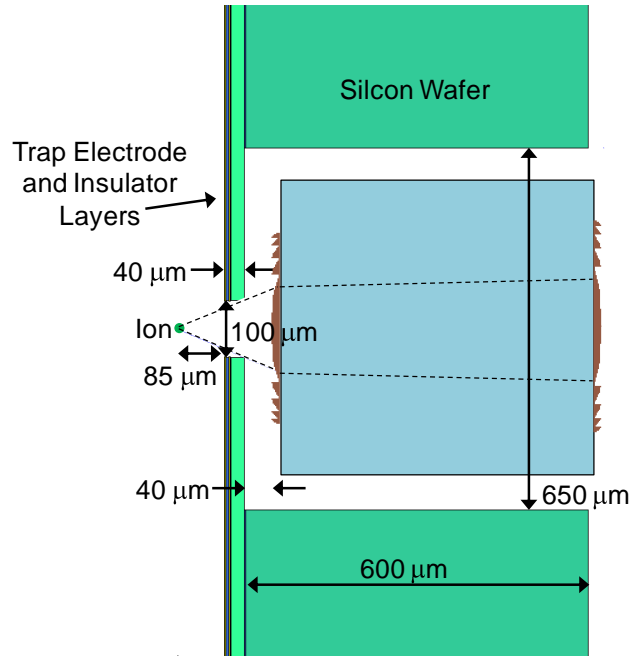

**Figure 2.** Section through the ion trap chip, showing representative dimensions. Dotted lines indicate the marginal rays from the ion passing through the trap slot, which is the aperture stop for the collection optics.

## 2.2. Installation in Vacuum Chamber

In addition to developing an appropriate optical design and integration scheme, we must also address several materials problems. All of the optical materials, materials for mechanical parts, and adhesives must be carefully considered and tested for compatibility with ultra-high vacuum and survivability throughout the required bake-out cycles to avoid contamination of the ion trap environment. We also need a multiple optical fiber vacuum feedthrough that does not compromise the integrity of the UHV chamber.

## 3. System Design

### 3.1. Integration Approach

Our method of integrating the optical system with the ion-trapping chip needs to accurately and rigidly connect the lens and fibers to the back side of the chip without interfering with the loading of ions in the trap. The configuration we designed is illustrated in the drawings of Figures 3 and 4. We use a microlens to efficiently collect light from the fluorescing ion and couple this light into a multimode fiber optic, the output of which is directed to a photomultiplier tube (PMT) outside of the chamber. We mount and polish the fiber in a commercial MT ferrule which, in addition to fixing the position of the fibers, provides a platform to which we mount the collection lens. The axial position of the lens relative to the ion and the fiber-end faces is determined by the geometry of a ceramic spacer placed between the ferrule and the lens. This subassembly is mated directly to the back surface of the chip. A stabilizer bracket is added to reduce stress at the chip-spacer interface due to the relatively large moment arm of the fibers and ferrule.

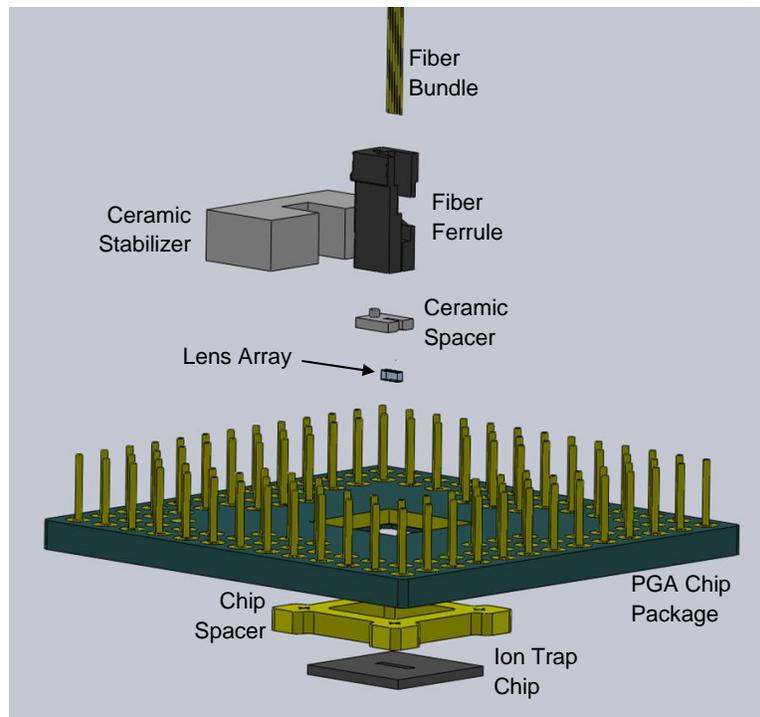

**Figure 3.** Exploded assembly drawing of the packaged ion trap and the integrated optics.

The ferrule, as purchased from the manufacturer, includes positions for 12 fibers on a 250 µm pitch. The full 6.4 mm width of the ferrule would block the entire loading slot of the chip, so it was necessary to cut the ferrule in half. The resulting ferrule has 6 fibers positions and only blocks part of the loading slot.

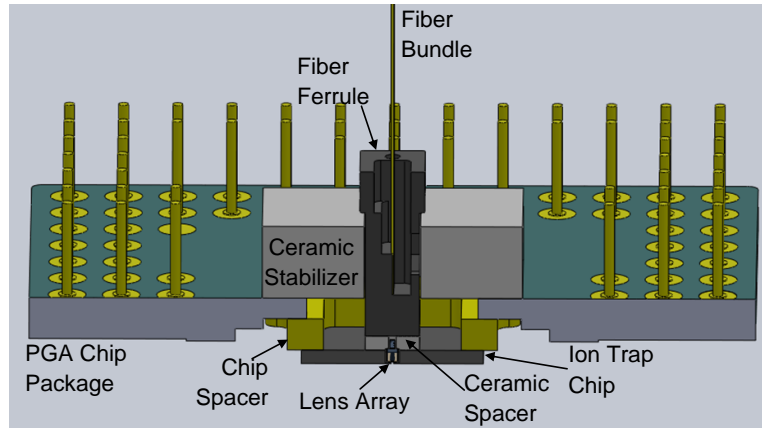

**Figure 4.** Assembly drawing of the optics integrated with the trap chip.

*3.2. Lens Design, Performance, and Trap-specific Features*

We chose a lens design form employing two diffractive surfaces so that optical power would be distributed between the two surfaces. This results in larger minimum feature sizes in the diffractive elements, and consequently greater efficiency. Further, the two surfaces allows for more degrees of freedom in the design, resulting in better aberration performance and looser tolerances. A ray trace drawing of this design is shown in Figure 5. Even with the power distributed, the two surfaces are quite fast: the first surface is f/1.44 and the second is f/1.72. The diffraction efficiencies of the surfaces were computed using rigorous coupled-wave theory, assuming that the lenses could locally be approximated by a linear grating [14], [15]. For an eight-level binary diffractive optic, the predicted efficiencies are 69.7% and 79.6%, or 55.5% for the cascaded elements. The lens material is UV-grade fused silica, which has high transmission at the design wavelength of 397 nm and is compatible with the UHV environment.

The nominal geometrical fiber coupling performance of the two-surface optic is excellent. All rays traced at each of the different ion positions considered are accepted by the fiber, implying 100% nominal fiber coupling efficiency (not including diffraction efficiency and reflection losses.) The ray trace model indicates that the image is diffraction limited on-axis and at ion positions up to 10 μm off-axis. The diffraction limited spot diameter is 7.8 μm. At the 10 μm off-axis point, the image spot is moved by 30 μm, which is still within the fiber core (specifications of the fiber are discussed in Sec. 3.3 below.)

A gold mask surrounds the lens apertures on the ion-facing surface, which limits the dielectric area within the field of view of the ion, mitigating charging effects. In the fully assembled device this mask is connected electrically to the chip's ground plane, providing a dissipating path for charge build-up on illumination.
The optical design is described in detail in Ref. [16].

*3.3. Fibers and Ferrule*

We chose Polymicro's FVP100110125 fiber for both its optical properties and its coating's UHV compatibility and temperature range [17]. The key parameters of this fiber are its core diameter of 100 μm, outer diameter of 124 μm ± 3 μm, and its numerical aperture of 0.22, which corresponds to an acceptance cone with a full-angle of 24.5 degrees. In addition, this fiber has low attenuation at 397 nm. The choice of a large core diameter, large numerical aperture fiber allows for loose positioning tolerances. The fiber core is glass and the buffer, or outer coating, is polyimide. Both of these materials are compatible with UHV and high temperatures.

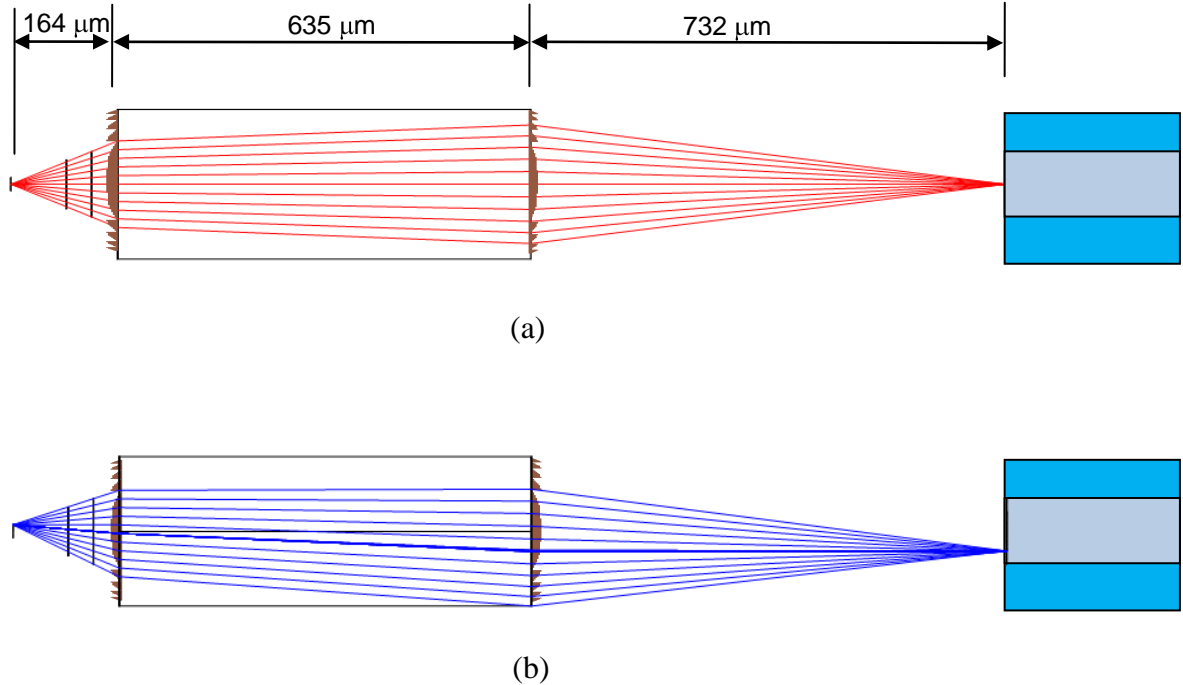

**Figure 5.** Ray trace depictions of a lens design using two diffractive surfaces to couple the light from the ion into a fiber. The first surface is f/1.44 and the second is f/1.72. (a) Ion is on-axis. (b) Ion is 10 μm off-axis.

The MT ferrule which holds the fiber is made of a glass-filled epoxy, manufactured by USConec [18]. The manufacturer did not provide data on the material's vacuum and temperature performance above 80 °C, so we tested this component in conjunction with the adhesives as described in Sec. 3.5.

*3.4. Mechanical Parts*

The spacer between the lens and the fiber ferrule and the mechanical stabilizer are machined from Macor ceramic. This material is highly compatible with the UHV environment, and has a coefficient of thermal expansion that is compatible with the glass-epoxy ferrule, fused silica lens, and silicon trap chip.

*3.5. Adhesives and UHV Testing*

We require an adhesive epoxy to pot the fibers in the ferrule and to mate together the ferrule, spacer, lens, and chip. We evaluated several low outgassing adhesives by building test assemblies of fibers set in the ferrule, and with the ferrule face attached to a metal surface. 3M's Scotch-Weld Epoxy Adhesive 2216 B/A [19] survived temperatures up to 178 °C, the highest of the adhesives tested. The glass-filled epoxy material of the ferrule held up under these high temperature conditions as well. We then verified that the adhesive and the fiber ferrule did not outgas significantly. This was done by performing a bake-out and evacuation cycle at 160 °C over 5 days, which yielded pressures in the $10^{-11}$ Torr range with no damage to the ferrule or adhesive. Based on these results, we selected the 2216 to bond our optical components together and affix the fibers in the ferrule. To provide a safety margin, we do not bake our chamber at temperatures higher than 150 °C when epoxies are present.

The current trap chip uses Johnson Matthey's JM7000 Silver Filled Cyanate Ester Die Attach Material in its packaging, so the high-temperature survivability and vacuum compatibility has already been demonstrated [20]. Since it is electrically conductive, we use a small bead of it to provide continuity between the gold mask of the lenses and the grounded, metalized backside of the chip.

For potential future use at cryogenic temperatures [21], we also submitted the 2216 epoxy and the ferrule to temperatures of approximately 10 K in a closed-cycle helium refrigerator for several hours and the bond did not fail.

*3.6. Vacuum Feedthrough*

We considered three types of vacuum feedthroughs to safely pass our optical fibers through the vacuum/air boundary: compression-type, pre-fabricated metal and glass, and custom-drilled with an epoxy seal. Each has benefits and drawbacks, but we ultimately chose the custom-drilled epoxy-sealed feedthrough as the lowest loss and most versatile choice.

Our feedthrough is made by drilling 250 μm holes in a vacuum blank using electrical discharge machining; Figure 6 shows a drawing of the part. The 124 μm diameter fibers are threaded through these holes and are sealed with our verified UHV-compatible epoxy, 3M's ScotchWeld 2216. This method accommodates multiple fibers and does not add transmission loss or apply stress to the fibers, but has the disadvantage that the fiber lengths inside the chamber cannot be adjusted once the epoxy is cured.

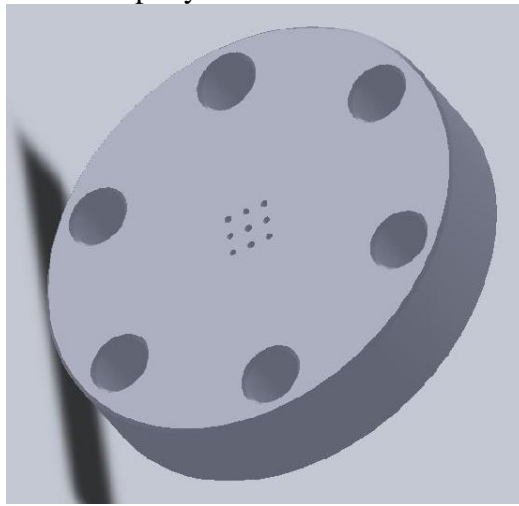

**Figure 6.** Custom-drilled 1.33" CF flange fiber feedthrough.

*Integration of fluorescence collection optics with a microfabricated surface electrode ion trap*9*Integration of fluorescence collection optics with a microfabricated surface electrode ion trap*   9

## 4. Assembly

### 4.1. Fabrication of Lenses

The fabrication process was similar to that used for most binary diffractive optics [22], [23]. In order to realize the optics in fused silica, the phase profiles optimized using the Zemax optical design package were approximated in the diffractive elements using eight discrete phase levels. To fabricate this surface relief required three separate electron beam write steps with each followed by a reactive ion etch to transfer the pattern into the substrate material. The smallest features that were written were approximately 120 nm and the desired maximum etch depth was 738 nm.

Linear arrays of five diffractive optics were placed on multiple 0.5 mm by 1.5 mm dies with a 250 μm pitch. The lens apertures were chosen to be square to maximize the collection area of each, with the apertures on the ion-facing surface being 140 μm squares and 250 μm squares on the reverse side. A sketch of the arrangement of the front surface of the chip is shown in Figure 7. Individual dies were cut from the parent wafers using a dicing saw.

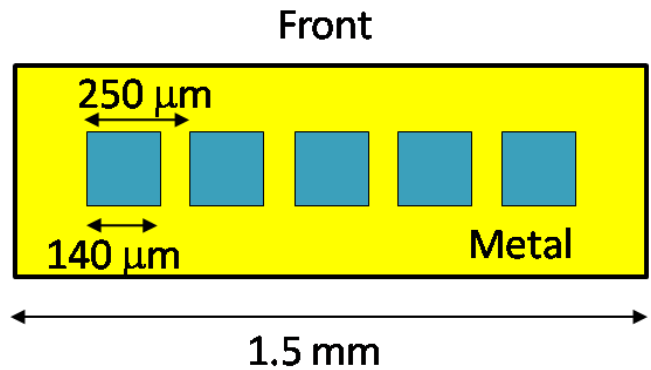

**Figure 7.** Front view of the optics die showing the five diffractive elements and the gold mask.

### 4.2. Assembly and Integration of Optics and Chip

4.2.1. Alignment Setup

When aligning the optics to the trap chip, the size and placement of the optics made it impossible to align them visually or using fiducial marks. As well, most of the surfaces are not suitable for mechanically registering parts. This required us to construct an alignment arrangement where the components are referenced against the position of optical beams simulating an ion and actively optimizing the throughput of the system. These "ion" sources were created using an array of 12 single mode fibers where light has been coupled into two selected fibers, allowing us to align the optics in all six degrees of freedom. These reference sources were placed on a fixed platform, as shown in the photograph of Figure 8.

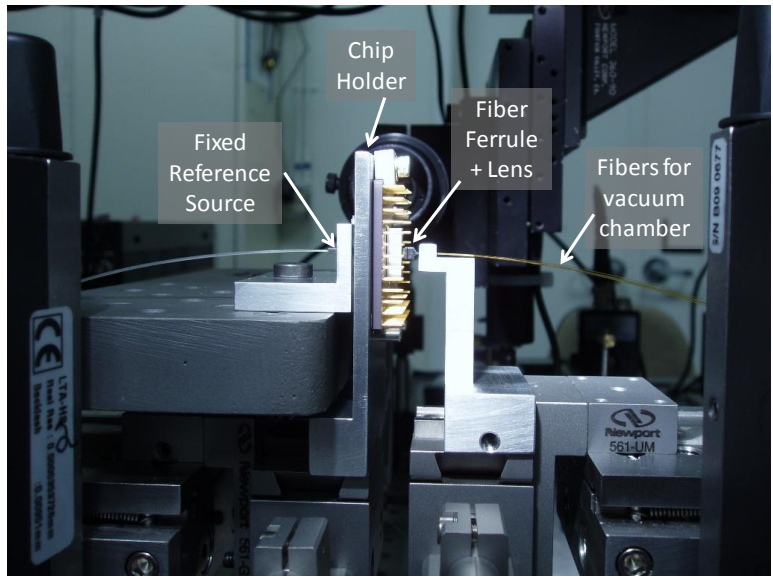

**Figure 8.** Photograph of custom fixtures for holding fixed reference source, chip and fiber ferrule.

4.2.2. Alignment and Assembly Process

The optics are assembled and integrated with the trap chip in several delicate steps, described in Ref. [16]. First, the ceramic spacer is attached to the end of the ferrule. Second, the multimode-fiber ferrule is aligned laterally to the reference sources. Third, the lens is aligned laterally to the reference sources and fibers and affixed to the spacer, where the spacer thickness sets the axial position of the lens. The result of this is shown in Figure 9. Next, the trap chip is aligned to the reference sources and the already aligned lens/spacer/fiber subassembly is mated to the back side of the chip. Finally, the ceramic stabilizer is added around the ferrule for additional support; the resulting assembly is shown in Figure 10.

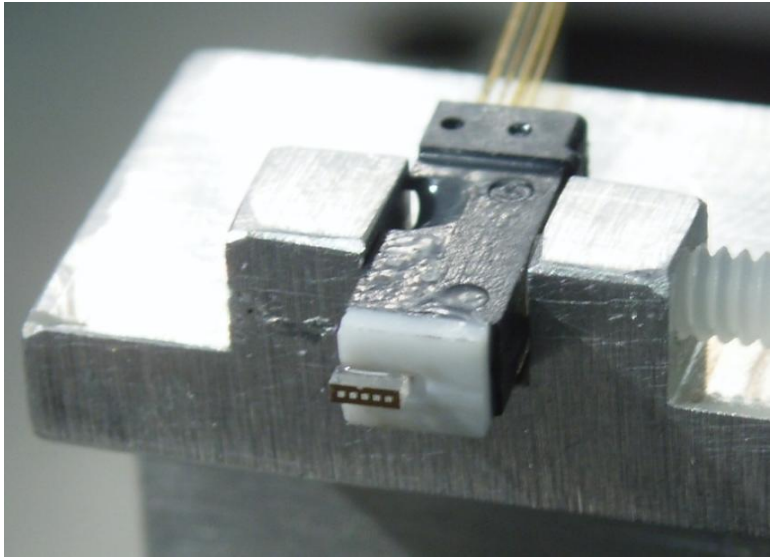

**Figure 9.** Lens array mounted on the spacer, with the epoxy cured. The gold mask and the five lens apertures are clearly visible.

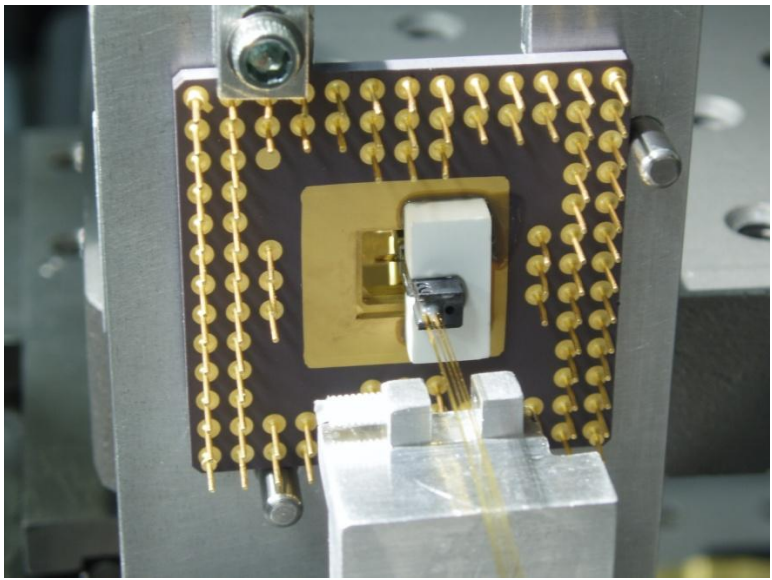

**Figure 10.** Completed ion trap with optics integrated.

*4.3. Introduction of Optics-Chip Assembly Into Vacuum Chamber*

The atomic oven that serves as a source of $Ca^+$ ions, consisting of a ceramic tube filled with calcium that is wrapped with a tungsten heating wire, was previously located behind the center of the chip to avoid trap electrode contamination [20]. With the optics in place, it must be shifted slightly to one side because the optics obstruct approximately half of the loading slot. We employed a mock-up chip with a dummy optical system while installing the oven to position it with the appropriate offset. This also allowed us to verify that the optical system would not be in thermal contact with the oven without risking damage to the functional fiber/chip assembly. Figure 11 shows the position of the oven relative to the optics and the chip.

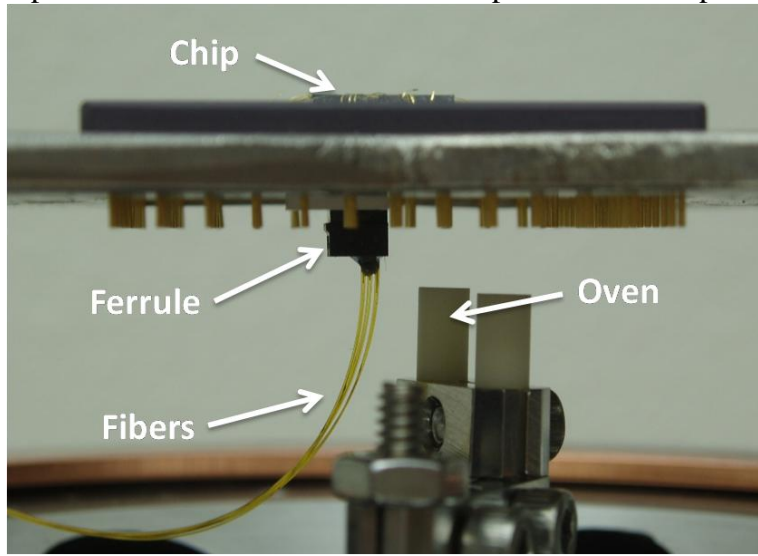

**Figure 11.** Detail of fiber/chip assembly in chamber.

Once the oven position was fixed, we installed the functional fiber/chip assembly and sealed the optical feedthrough. The vacuum chamber is built from stainless steel components with ConFlat flanges, and we evacuated and baked the chamber at 150 °C for 12 days while pumping at a rate of >150 l/s. After the chamber cooled, the pressure reached $1.6 \times 10^{-11}$ Torr, maintained with a 20 l/s ion pump. This demonstrates that our components do not contribute excess contamination and our vacuum feedthrough is effectively sealed.

## 5. Trapping and Results

After the bake-out and evacuation cycle, we moved the trap chamber to an optical table and configured our trapping system as in previous work [20]. We trapped successfully, loading the ion near the center of the linear ion trap (two electrodes away from the exact trap center). The geometry of the ion trap is shown in Figure 12. We used the same control voltage solution that has been successfully used with identically-fabricated trap chips lacking integrated optics. We shuttled the ion back and forth across the chip 30,000 times without loss over a linear section of 770 μm with a shuttling time of 1 ms, traversing the region of the optics on each pass and demonstrating that the dielectric surface of our lens close to the ion does not preclude trapping in that region. Further, we do not observe any pertubation in the ion's nominal position relative to that observed using an identical trap without integrated optics, within the capabilities of our measuring system. During the fast shuttling of the ion over the optics, the 397 nm laser was only located at the loading region. In addition, we shuttled and held the ions directly above a lens, with the laser following, and observed trapped ion lifetimes on the same time scale (tens of minutes to hours) as observed in the loading region of this trap. Here, we successfully detected fluorescence photons emitted from the ion and coupled through the fiber at the output facet of the fiber. We will characterize and quantify this, including collection efficiencies and crosstalk between channels, in a future publication.

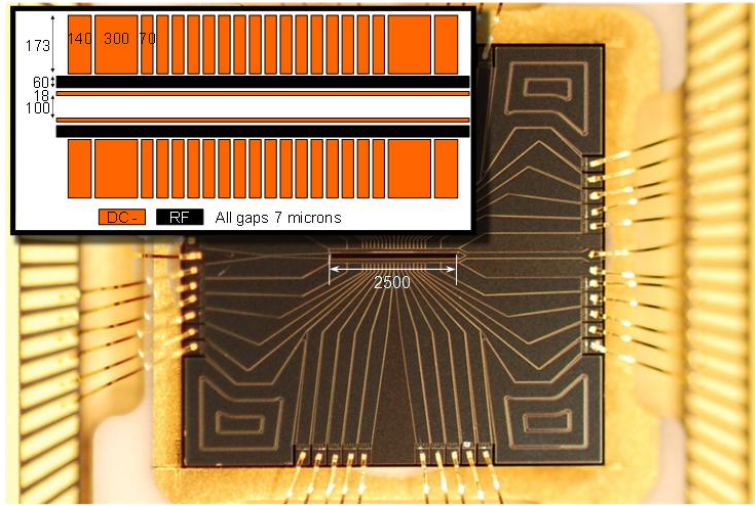

**Figure 12.** Photograph and drawing (inset) showing the geometry of the ion-trapping chip. All dimensions are in micrometers.

## 6. Conclusion

We have successfully integrated optics that couple the fluorescence from a trapped ion into an optical fiber with a working ion trap system. Careful optical design work and analysis work was performed to identify a practical design form that could be integrated with an ion-trapping chip, incorporating a number of design considerations unique to the trap. This optical element was successfully fabricated using binary optics techniques. These optics were aligned and integrated with the ion-trapping chip and optical fibers using a commercial fiber ferrule, custom ceramic components, and adhesives. All components and adhesives were qualified for use at UHV, including the necessary bake-out cycle. The optics-trap assembly was placed in a vacuum chamber and a pressure of $1.6 \times 10^{-11}$ Torr was achieved. Ions were successfully trapped and shuttled across the region where the optics were present several thousand times. Finally, we have successfully detected fluorescence from the ion using the integrated optics.

Future work will completely characterize the efficiency of the optics and their robustness to misalignments. In addition, more efficient optics with a continuous surface relief or a smaller number of surfaces will be investigated. Future designs incorporating both excitation and collection optics should also be considered.

## Acknowledgements

The authors would like to acknowledge the Steane/Lucas group at the University of Oxford, the Blatt group at Universität Innsbruck, D. Wineland's group at NIST Boulder, and D. Kielpinski at Griffith University.

Sandia is a multiprogram laboratory operated by Sandia Corporation, a Lockheed Martin Company, for the United States Department of Energy's National Nuclear Security Administration under Contract DE-AC04-94AL85000.